\documentclass[showpacs,twocolumn,preprintnumbers,superscriptaddress,amsmath,amssymb,nofootinbib]{revtex4}

\input psfig.sty

\usepackage[dvips]{graphicx}
\usepackage{color}
\usepackage{latexsym}
\usepackage{amsmath}
\usepackage{amssymb}
\usepackage{eufrak}
\usepackage{euscript}
\usepackage{pstricks}
\usepackage{graphics}
\usepackage{graphicx}
\usepackage{picture}
\def\[{\left\lbrack}
\def\]{\right\rbrack}

\def\({\left(}
\def\){\right)}

\newcommand{\bee}{\begin{equation}}
\newcommand{\eee}{\end{equation}}
\newcommand{\eaa}{\end{eqnarray}}
\newcommand{\baa}{\begin{eqnarray}}

\def\ni{\noindent}

\def\no{\nonumber}

\begin{document}

\title{\Large Dark energy models through nonextensive Tsallis' statistics}

\author{Rafael da C. Nunes}\email{rafaelnunes@uern.br}
\author{Ed\'esio M. Barboza Jr.} \email{edesiobarboza@uern.br}

\affiliation{Departamento de F\'isica, Universidade do Estado do Rio Grande do Norte, 59610-210, Mossor\'o - RN, Brasil}
\author{Everton M. C. Abreu}\email{evertonabreu@ufrrj.br}
\affiliation{Grupo de F\' isica Te\'orica e Matem\'atica F\' isica, Departamento de F\'{i}sica, Universidade Federal Rural do Rio de Janeiro, 23890-971, Serop\'edica - RJ, Brasil}
\affiliation{Departamento de F\'{i}sica, Universidade Federal de Juiz de Fora, 36036-330, Juiz de Fora - MG, Brasil}
\author{Jorge Ananias Neto}\email{jorge@fisica.ufjf.br}
\affiliation{Departamento de F\'{i}sica, Universidade Federal de Juiz de Fora, 36036-330, Juiz de Fora - MG, Brasil}
\date{\today}

\begin{abstract}
\noindent The accelerated expansion of the Universe is one of the greatest challenges of modern physics.  One candidate to explain this phenomenon is a new field called dark energy.  In this work we have used the Tsallis nonextensive statistical formulation of the Friedmann equation to explore the Barboza-Alcaniz and Chevalier-Polarski-Linder parametric dark energy models and the Wang-Meng and Dalal vacuum decay models.  After that, we have discussed the observational tests and the constraints concerning the Tsallis nonextensive parameter.
\end{abstract}

%\pacs{98.80.Jk; 98.80.-k; 04.20.Cv}

\maketitle

\section{Introduction}

The discovery that the universe is expanding at an accelerated rate has given rise to several hypotheses and speculations about the gravity theory and the material content of the Universe. The cosmological constant, the ``Einstein's big mistake", now reinterpreted as the density of energy associated to the quantum vacuum, is back to the game as the strongest candidate to explain such unexpected fact. However, despite its excellent agreement with the majority of cosmological data, its value required by the observations is far from the value predicted by the field theory by at least $60$ orders of magnitude \cite{weinberg}. This astounding disagreement between theory and observation has led the physicists to explore other routes to explain the acceleration of the universe. Beyond the cosmological constant, the simplest assumption that keeps the general relativity untouched is that the universe is pervading by a fluid with a negative pressure in order to violate the strong energy condition, that is, $\rho+3p<0$. This fluid is called dark energy (DE) and it is characterized by the equation of state (EoS) parameter $w=p/\rho$. Unlike the dark matter (DM), whose the existence was already confirmed, but its nature remains unknown, the DE existence itself has not yet been confirmed.   Since the DM and DE origins remain unknown, a coupling in the dark sector can not be discarded and it constitutes another appealing hypothesis to alleviate the cosmological constant problem.  In this paper, we will use the nonextensive Tsallis' statistics in order to investigate the 
Barboza-Alcaniz \cite{barboza} and Chevalier-Polarski-Linder \cite{CPL} time-dependent parametric models of DE and the Wang-Meng \cite{wang-meng} and Dalal \cite{dalal} vacuum decay models.

There are theoretical evidences that the understanding of gravity has been greatly benefited from a possible connection with thermodynamics. Pioneering works of Bekenstein \cite{BEK} and Hawking \cite{HAW} have described this issue. For example, quantities as area and mass of black-holes are associated with entropy and temperature respectively. Working on this subject, Jacobson \cite{Jac} interpreted Einstein field equations as a thermodynamic identity. Padmanabhan \cite{PAD} gave an interpretation of gravity as an equipartition theorem. 

Recently, Verlinde \cite{verlinde} brought an heuristic derivation of gravity, both Newtonian and relativistic, at least for static spacetime.  The equipartition law of energy has also played an important role. The analysis of the dynamics of an inflationary Universe ruled out by the entropic gravity concept was investigated in \cite{inflation}.
On the other hand, one can ask: what is the point of view of gravitational models coupled with thermostatistical theories and vice-versa?  

The concept introduced by Verlinde is analogous to Jacobson's \cite{Jac} one, who proposed a thermodynamic derivation of Einstein's equations.  The result has shown that the gravitation law derived by Newton can be interpreted as an entropic force originated by perturbations in the information ``manifold" caused by the motion of a massive body when it moves 
away from the holographic screen. An holographic screen can be understood as a storage device for information which is constituted  by bits. Bits are the smallest unit of information. Verlinde used this idea  together with the Unruh result \cite{unruh} and he obtained Newton's second law. Moreover, assuming  the holographic principle together with the equipartition law of energy,  the Newton law of gravitation could be derived.  The connection between nonextensive statistical theory and the entropic gravity models \cite{ananias,abreu} make us to realize an arguably bridge between nonextensivity and gravity theories.  More specifically, theories which consider accelerated models with DE approaches.
In this way, the use of the equipartition energy statistical law can lead us to imagine the role of alternative statistical theories beyond Boltzmann-Gibbs standard theory.  
It was constructed some years ago an extension of the usual Boltzmann-Gibbs theory (BG) that is called 
Tsallis  thermostatistics (TT) formalism \cite{tsallis,tsallis2}. To sum up, this formalism always considers the entropy formula as an extensive  quantity and has been successfully applied in many physical models. 

We have organized our paper in the following way: in section \ref{vf} we will provide a brief review of Tsallis' approach. In section \ref{Res} we will demonstrate that the NE version of the Friedmann equation can be written as a function of the radiation, matter and DE densities.  A new normalization condition, as a function of the $q$-parameter will be obtained.
In section \ref{FET} we will obtain new values for $q$-parameter for different DE models.  Specifically, the Barboza-Alcaniz and Chevalier-Polarski-Linder parametric models.  In section V we will analyze nonextensively, two vacuum decay models, the Wang-Meng and the Dalal models and the observational constraints will be discussed in section VI. The conclusions will be depicted in the last section.

\section{Tsallis' statistics in a nutshell}
\label{vf}

One of the reasons that the study of entropy has been an interesting task through the years is the fact that it can be considered as a measure of information loss concerning the microscopic degrees of freedom of a physical system when depicting it in terms of macroscopic variables.  Appearing in different scenarios, entropy can be deemed as a consequence of gravitational framework \cite{nicolini}.  These issues motivated some of us to consider other alternatives to the standard BG theory in order to work with Verlinde's ideas together with other subjects \cite{abreu}.

Tsallis \cite{tsallis} has proposed an important extension of the Boltzmann-Gibbs statistical theory and curiously, in a technical terminology, this model is also currently referred to as nonextensive statistical mechanics (NE). TT formalism defines a nonadditive entropy given by
\begin{eqnarray}
\label{nes}
S_q =  k_B \, \frac{1 - \sum_{i=1}^W p_i^q}{q-1}\;\;\;\;\;\;\qquad (\sum_{i=1}^W p_i = 1)\,\,,
\end{eqnarray}
where $p_i$ is the probability of the system to be in a microstate, $W$ is the total number of configurations and 
$q$, known in the current literature as Tsallis parameter or nonextensive (NE)  parameter, is a real parameter which quantifies the degree of nonextensivity. 
The definition of entropy in TT formalism motivated the study of multifractals systems and it also possesses the usual properties of positivity, equiprobability, concavity and irreversibility.
It is important to stress that Tsallis formalism contains the Boltzmann-Gibbs statistics as a particular case in the limit $ q \rightarrow 1$ where the usual additivity of entropy is recovered. Plastino and Lima \cite{PL}  
have derived a NE equipartition law of energy whose expression can be written as

\begin{eqnarray}
\label{ge}
E = \frac{1}{5 - 3 q} N k_B T\,\,,
\end{eqnarray}
where the range of $q$ is $ 0 \le q < 5/3 $.  For $ q=5/3$ (critical value) the expression of the equipartition law of energy, Eq. (\ref{ge}), diverges. It is also easy to observe that for $ q = 1$,  the classical equipartition theorem for each microscopic degrees of freedom can be recovered. 

As an application of NE equipartition theorem in Verlinde's formalism we can
use the NE equipartition formula, i.e., Eq. (\ref{ge}).  Hence, we can obtain a modified acceleration formula given by \cite{abreu}
\begin{eqnarray}
\label{accm}
a = G_{NE} \, \frac{ M}{r^2},
\end{eqnarray}
where $G_{NE}$ is an effective gravitational constant which is written as

\bee
\label{S}
G_{NE}=\,\frac{5-3q}{2}\,G\,\,.
\eee
From result (\ref{S}) we can observe that the effective gravitational constant depends on the NE parameter $q$. For example, when $q=1$ we have $ G_{NE}=G$ (BG scenario) and for $q\,=\,5 / 3$ we have the curious and hypothetical result which is $G_{NE}=0$.  This result shows us that $q\,=\,5/3$ is an upper bound limit when we are dealing with the holographic screen.  Notice that this approach is different from the one demonstrated in \cite{cn,kok}, where the authors considered in their model that the number of states is proportional to the volume and not to the area of the holographic screen.

\section{FRW cosmologies from NE Tsallis' statistics}
\label{Res}

It was demonstrated in \cite{abreu} that one modification in the dynamics of the Friedmann-Robertson-Walker (FRW) Universe in NE Tsallis' statistics can be obtained simply by making the prescription $G\to G_{NE}=(5-3q)G/2$ in the standard field equations. Thus, the equations of motion in the NE statistics are
\begin{equation}
\label{Friedmann_eq}
H^2+\frac{k}{a^2}=\frac{4(5-3q)\pi G}{3}\rho\,\,,
\end{equation}

\noindent and

\begin{equation}
\label{field_eq2}
2\frac{\ddot{a}}{a}+H^2+\frac{k}{a^2}=-\frac{4(5-3q)\pi G}{3}p\,\,,
\end{equation}

\noindent where $H=\dot{a}/a$ is the Hubble function and $\rho$ and $p$ are, respectively, the total density and pressure of the fluid. These equations can be combined to obtain the conservation equation,

\begin{equation}
\label{conservation}
\dot{\rho}+3H(\rho+p)=0\,\,.
\end{equation}

\noindent For a Universe filled by radiation, matter (baryonic plus DM) and DE, the Friedmann equation (\ref{Friedmann_eq}) becomes

\begin{eqnarray}
\label{Friedmann_eq2}
\frac{H^2}{H_0^2}=\frac{5-3q}{2}\Big[\frac{\Omega_{\gamma,0}}{a^{4}}+\frac{\Omega_{m,0}}{a^{3}}+\frac{\Omega_{k,0}}{a^{2}}+\Omega_{x,0}f(a)\Big]\,\,,
\end{eqnarray}

\noindent where 

\begin{equation}
\label{DE_density_function}
f(a)\equiv\frac{\rho_x}{\rho_{x,0}}=a^{-3}\exp\Big(-3\int_1^a\frac{w(a')da'}{a'}\Big)\,\,,
\end{equation}

\noindent and the subscript ``$0$" denotes the present time value of a quantity; $\Omega_{i,0}=8\pi G\rho_{i,0}/(3H_0^2)$ is the density parameter of the {\it i-th} component ($i=\gamma,\, m,\,\mbox{and}\,x$ for radiation, matter and DE, respectively); $\Omega_{k,0}=-k/H_0^2$ is the curvature density parameter and $w(a)=p_x/\rho_x$ is the EoS parameter of DE which we assume that it is a function of time. Here we have used the convention $a_0=1$. In the NE scenario, the normalization condition reads as

\begin{equation}
\label{normalization}
\Omega_{\gamma,0}+\Omega_{m,0}+\Omega_{k,0}+\Omega_{x,0}=\frac{2}{5-3q}\,\,,
\end{equation}

\ni which is an interesting result since it can show us, one more time, that the value $q=1$ recovers the standard normalization condition.   Values of $q > 5/3$ which brings a negative normalization condition, makes no sense.  Another consequence of the above equation is that at least one of the $\Omega$-densities will be a function of $q$.  This result was obtained in \cite{abreu} but not for $\Omega_{x,0}$.  So, from (\ref{normalization}), we can, alternatively calculate the $q$-parameter as a function of the integral in (\ref{DE_density_function}), since we have the value of the other three densities in (\ref{normalization}), of course.  New values for $q$-parameter for different models will be obtained in the next section.

\section{Dark energy models}
\label{FET}

The investigation of DE models brings new analysis not only in cosmology but also in high energy physics.  With these goals in mind, in this section we will discuss two recent parameterization for DE EoS.  Later, on section VI, we will connect these ideas with the nonextensive approach.

Our first model concerning DE is the Barboza-Alcaniz (BA) parametric model firstly studied in \cite{barboza}. The second one is the Chevalier-Polarski-Linder (CPL) parametric model \cite{CPL}, one of most studied in the literature.

\subsection{Barboza - Alcaniz parameterization}

The Barboza-Alcaniz (BA) parameterization is given by

\begin{equation}
\label{wa_BA}
w(a)=w_0+w'_0 \frac{1-a}{1-2a+2a^2}\,\,,
\end{equation}

\noindent or, in terms of the redshift $a=(1+z)^{-1}$

\begin{equation}
\label{wz_BA}
w(z)=w_0+w'_0 \frac{z(1+z)}{1+z^2}\,\,.
\end{equation}

\noindent In the above equations $$\left. w'_0=\frac{dw}{dz}\right|_{z=0}\,\,,$$ is a parameter that measures the EoS time dependence. For this parameterization, the density function is

\begin{equation}
\label{BA_fz}
f(z)=(1+z)^{3(1+w_0)}(1+z^2)^{3w'_0/2}\,\,.
\end{equation}

The main characteristic of the EoS parameterization (\ref{wz_BA}) is that it is a well behaved function of the redshift during the whole history of the universe ($z\in[-1,\infty[$), which allows one to introduce in its functional form the important case of a quintessence scalar field ($-1<w(z)<1$).  By noting that $w(z)$ has absolute extremes in $z_{\pm}=1\pm\sqrt{2}$ corresponding, respectively, to $w_-=w(z_-)=w_0-0.21w'_0$ and $w_+=w(z_+)=w_0+1.21w'_0$ it is possible to separate the parameter space $(w_0,w'_0)$ into defined regions associated to distinct DE models which can be confronted with the observational constraints to confirm or rule out a given DE model. For $w'_0>0$, $w_-$ is a minimum and $w_+$ is a maximum and for $w'_0<0$ this is inverted. Since for quintessence and phantom scalar fields the EoS is constrained by $-1\leq w(z)\leq1$ and $w(z)<-1$, respectively, the region occupied in the $(w_0,w'_0)$ plane by these fields can be determined easily. To discuss quintessence, we have that $-1\leq w_0-0.21w'_0$ and $w_0+1.21w'_0\leq1$ if $w'_0>0$ and $-1\leq w_0+1.21w'_0$ and $w_0-0.21w'_0\leq1$ if $w'_0<0$. Concerning phantom fields, we can write $w'_0<-(1+w_0)/1.21$ if $w'_0>0$ and $w'_0>(1+w_0)/0.21$ if $w'_0<0$. Points out of these bounds correspond to mixed DE scenarios that crossed or will cross the phantom separation line.

\subsection{Chevalier - Polarski - Linder parameterization}

The Chevalier-Polarski-Linder (CPL) parameterization is given by 

\begin{equation}
\label{CPL_a}
w(a)=w_0+w'_0(1-a)\,\,,
\end{equation}

\noindent or equivalently 

\begin{equation}
\label{CPL_z}
 w(z)=w_0+w'_0\frac{z}{1+z}\,\,.
\end{equation}

\noindent Now, the density function is

\begin{equation}
\label{CPL_fz}
f(z)=(1+z)^{3(1+w_0+w'_0)}\exp\Big[-3w'_0\frac{z}{1+z}\Big]\,\,.
\end{equation}

\noindent The CPL parameterization has an absolute extreme in $w_{\infty} = w(z \to \infty) = w_0 + w'_0$. For $w'_0 > 0, w_{\infty}$ is a maximum whereas for $w'_0 < 0$ it is a minimum. Thus, the region occupied by phantom fields is determined by the constraints $w_{\infty} < -1$ and $w'_0 > 0$, whereas similar
constraints cannot be obtained for the quintessence case.  Note that for the CPL parameterization, when $z\to -1 (a \to\infty)$, $f(z)$ explodes if $w'_0 > 0$ while $f(z)$, given by Eq. (\ref{BA_fz}), blows up in this limit if
$w_0 < -1$. Thus, the roles of the parameters $w_0$ and $w'_0$ are interchanged in this limit, in the sense that while for BA parameterization the fate of the Universe will be dictated by the equilibrium part $w_0$, for the CPL parameterization the future of the Universe will be driven by the time-dependent term $w'_0$.

\section{Vacuum decay models}

Beyond DE parametric models, concerning the cosmological constant problem, another line of attack is to assume that the cosmological term evolves with  time. Here, we will investigate the consequences of the nonextensive Tsallis' statistics on two vacuum decay models: 
the Wang-Meng \cite{wang-meng} and the Dalal \cite{dalal} models. In this case, the conservation equation (\ref{conservation}) becomes

\begin{equation}
\label{conservation_coupled}
\dot{\rho}_{dm}+3H\rho_{dm}=-\dot{\rho}_{\Lambda}\,\,.
\end{equation}

\ni which connects the evolution in time of the DM and the cosmological constant terms.

\subsection{Wang-Meng Model}

If DE and DM interact, the energy density of this latter component will dilute at a different rate compared to its standard evolution, $\rho_{dm}\propto a^{-3}$. Thus, the deviation from the standard dilution may be characterized
by a constant $\epsilon$ that measures the deviation of the standard evolution law, such that

\begin{equation}
\label{wang_meng_law}
\rho_{dm}=\rho_{dm,0} a^{-3+\epsilon}\,\,.
\end{equation}

\noindent By substituting (\ref{wang_meng_law}) into (\ref{conservation_coupled}) and solving the resulting equation, we have that

\begin{equation}
\label{density_wang_meng}
\rho_{\Lambda}=\rho_{\Lambda,0}-\frac{\epsilon\rho_{dm,0}}{\epsilon-3}a^{-3+\epsilon}+\frac{\rho_{dm,0}\epsilon}{\epsilon-3}\,\,.
\end{equation}

\noindent Now, the Friedmann equation (\ref{Friedmann_eq2}) reads

\begin{eqnarray}
\label{Friedmann_wm}
\frac{H^2}{H_0^2}&=&\frac{5-3q}{2}\Big[\frac{\Omega_{\gamma,0}}{a^{4}}+\frac{\Omega_{b,0}}{a^{3}}+\Big(1-\frac{\epsilon}{\epsilon-3}\Big)\frac{\Omega_{dm,0}}{a^{3-\epsilon}}+\nonumber\\&+&\frac{\Omega_{k,0}}{a^{2}}+\Omega_{\Lambda,0}+\frac{\epsilon}{\epsilon-3}\Big]\,\,,
\end{eqnarray}

\noindent where the density parameters are related by (\ref{normalization}) with $\Omega_{m,0}=\Omega_{dm,0}+\Omega_{b,0}$ and $\Omega_{x,0}=\Omega_{\Lambda,0}$ ($\Omega_{b,0}=0.04$ \cite{planck} stands for the baryon density parameter).

\subsection{Dalal's model}

\noindent Another model proposed to alleviate the cosmological constant problem, assumes that the ratio between the dark components follow a power law \cite{dalal} which is

\begin{equation}
\label{dalal}
\rho_{\Lambda}=ra^{\xi}\rho_{dm}\,\,,
\end{equation}

\noindent where $r=\rho_{\Lambda,0}/\rho_{dm,0}$ and $\xi\ne0$ is an addimensional parameter that measures the coupling intensity. In this scenario, the $\Lambda$CDM model is recovered when $\xi=3$. Substituting (\ref{dalal}) into (\ref{conservation_coupled}) and solving the resulting equation, we obtain that

\begin{equation}
\label{dala_matter_law}
\rho_{dm}=\rho_{dm,0}a^{-3}\Big(\frac{1+r}{1+ra^{\xi}}\Big)^{1-3/\xi}\,\,,
\end{equation}

\noindent and

\begin{equation}
\label{dala_vacuum_law}
\rho_{\Lambda}=\rho_{\Lambda,0}a^{-3+\xi}\Big(\frac{1+r}{1+ra^{\xi}}\Big)^{1-3/\xi}\,\,.
\end{equation}

\noindent Now, the Friedmann equation becomes

\begin{eqnarray}
\label{Friedmann_dalal}
\frac{H^2}{H_0^2}&=&\frac{5-3q}{2}\Big[\frac{\Omega_{\gamma,0}}{a^{4}}+\frac{\Omega_{b,0}}{a^{3}}+\frac{\Omega_{k,0}}{a^{2}} \\
&+&\Big(\Omega_{dm,0}+\Omega_{\Lambda,0}a^{\xi}\Big)a^{-3}\Big(\frac{1+r}{1+ra^{\xi}}\Big)^{1-3/\xi}\Big]\,\,.  \nonumber
\end{eqnarray}

\noindent As in the previous case, the density parameters are related to each other through (\ref{normalization}) by making the substitutions $\Omega_{m,0}=\Omega_{dm,0}+\Omega_{b,0}$ and $\Omega_{x,0}=\Omega_{\Lambda,0}$. Notice also that the nonextensivity parameter $q$ enters in the factor $r=\rho_{\Lambda,0}/\rho_{dm,0}$ by the normalization condition 
(\ref{normalization}). Motivated by the recent results of the CMB power spectrum\cite{cmb}, we will assume spatial flatness in the following analysis.

\section{Observational Constraints}

\begin{figure*}[t]
\centerline{\psfig{figure=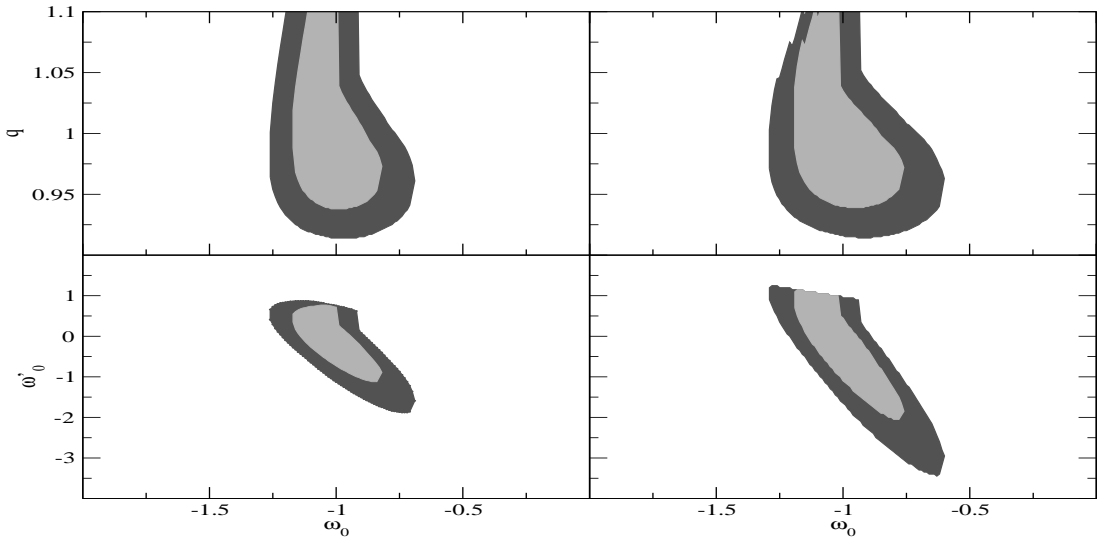,width=6.0truein,height=3.0truein,angle=270}}
\caption{The $\omega_0-q$ and $\omega_0-\omega'_0$ parametric spaces for BA (left) and CPL (right) parametric models. The contours are drawn for $\Delta\chi^2=2.30$ and 6.17.\label{fig1}}
\end{figure*}

\begin{figure*}[t]
\centerline{\psfig{figure=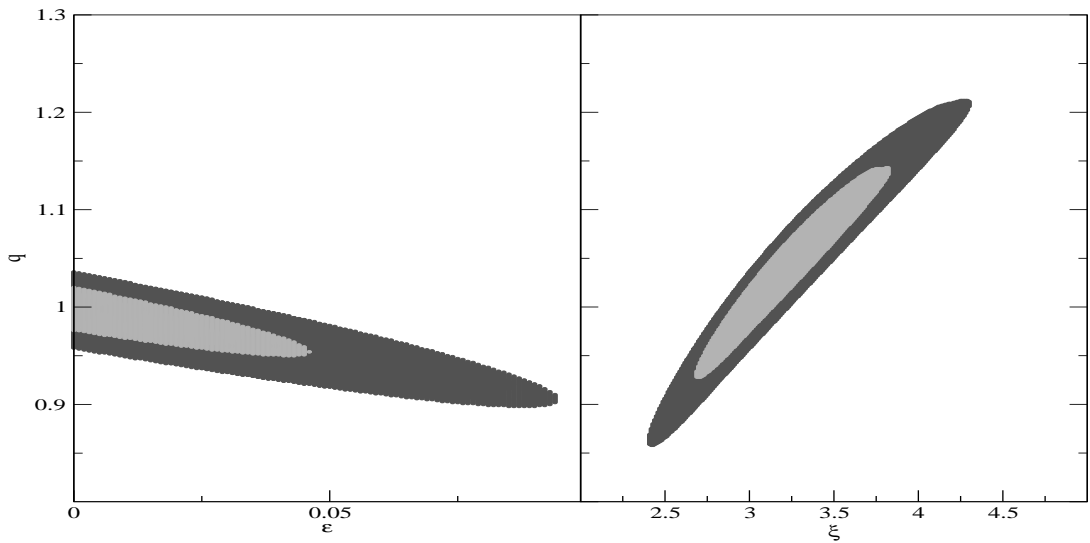,width=6.0truein,height=3.0truein,angle=270}}
\caption{The $\epsilon-q$ (left) and $\xi-q$ (right) parametric spaces. The contours are drawn for $\Delta\chi^2=2.30$ and 6.17. The best fit points are $(\epsilon,q,)=(0.00,1.02)$ and $(\xi,q,)=(3.18,1.03)$ with $\chi^2_{\nu}\equiv\chi^2/{\rm NDoF}=0.97$ for both scenarios.\label{fig2}}
\end{figure*}

\begin{table}[!h]
	\begin{center}
		\begin{tabular}{lccccc}
		\hline
		\hline\\
		Model& $\chi^2_{\nu}$&$ w_0$& $w'_0$& $ q$&$\Omega_{m,0}$\\
		\hline
		\hline\\
                   $\Lambda$CDM&0.96&$-1.00$&$-$&$0.99^{+0.02}_{-0.02}$&$0.28^{+0.02}_{-0.02}$\\
                   $w$CDM&0.97&$-1.04^{+0.10}_{-0.10}$&$-$&$0.98^{+0.04}_{-0.04}$&$0.28^{+0.02}_{-0.02}$\\
		BA&0.97&$-1.00^{+0.18}_{-0.17}$&$-0.26^{+0.98}_{-0.86}$&$0.97^{+0.13}_{-0.04}$&$0.29^{+0.01}_{-0.02}$\\
		CPL&0.97&$-0.98^{+0.22}_{-0.21}$&$-0.50^{+1.70}_{-1.55}$&$0.97^{+0.12}_{-0.04}$&$0.29^{+0.01}_{-0.02}$\\
		\hline
		\hline
		\end{tabular}
	\end{center}
	\caption{The best fit and the $1\sigma$ errors for one parameters for $\Lambda$CDM, $\omega$CDM, BA and CPL DE models.}
	\label{bf_de}
\end{table}

In order to discuss the current observational constraints of $w_{0}$,  $w'_{0}$ and the nonextensive parameter $q$, the Union 2.1 SN Ia sample of  Ref.~\cite{union}, which is an update of the Union 2 compilation and comprises 580 data points \cite{union}, will be used. We will also use the results of current BAO and CMB experiments to diminish the degeneracy between the parameters studied. For BAO measurements, the six estimates of the BAO parameter 
\begin{equation}
\label{BAO_parameter}
{\cal{A}} (z) = D_V{\sqrt{\Omega_{\rm{m}} H_0^2}},
\end{equation}
given in Table 3 of Ref.~\cite{blake} are used. In this latter expression, 
$D_V = [r^2(z_{\rm{BAO}}){z_{\rm{BAO}}}/{H(z_{\rm{BAO}})}]^{1/3}$
is the so-called dilation scale, defined  in terms of the dimensionless comoving distance $r$. For the CMB, only the
measurement of the CMB shift parameter~\cite{wmap}
\begin{equation}
{\cal{R}} = \Omega_{\rm{m}}^{1/2}r(z_{\rm{CMB}}) = 1.725 \pm 0.018\;,
\end{equation}
where $z_{\rm{CMB}} = 1089$ is used. Thus, in the present analysis, the function  
\begin{equation}
\chi^2 = \chi^{2}_{\rm{SNe}} + \chi^{2}_{\rm{BAO}} + \chi^{2}_{\rm{CMB}}\,\,,
\end{equation} 
which takes into account all the data sets mentioned above, is minimized.   Since we are interested only in the constraints over the DE parameters and in the nonextensitvity parameter, we have marginalized  the current value of the Hubble parameter, $H_0$.
% and the matter density parameter $\Omega_{m,0}$ are marginalized. 

Figure \ref{fig1} shows the results of the statistical analysis in $68\%$ and $95\%$ confidence levels. The left column figures shows the $q-w_{0}$ and $w_{0}-w'_{0}$ parameter spaces for the BA parameterization. The right column figures shows the $q-w_{0}$ and $w_{0}-w'_{0}$ parameter spaces for the CPL parameterization. We have marginalized $\Omega_{m,0}$ and we imposed the physical constraint $\omega_0+ \omega'_0<0$ upon both models in order to guarantee that the DE is subdominant at early times, i. e., $\rho_{DE}\ll\rho_{dm}$ for $z\gg1$. The best fit values  for these DE models are summarized in Table \ref{bf_de}. For the sake of comparison, we have also displayed the results for the $\Lambda$CDM and $w$CDM models.

Figure \ref{fig2} shows the $1\sigma$ and $2\sigma$ confidence regions for the nonextensivity parameter $q$ and the Wang-Meng coupling parameter $\epsilon$ (left) and the Dalal's coupling parameter $\xi$ (right). In both figures we have marginalized  $\Omega_{dm,0}$. According to the  thermodynamics constraints \cite{jj} we have used that $\epsilon>0$. The best fit values for the vacuum decay models are summarized in Table \ref{bf_vd}.

It is worth mention that, although our results for both scenarios, DE and vacuum decay, favor the BG statistics ($q=1$), there is enough space to nonextensive Tsallis' statistics, i. e., $q\ne1$.

\begin{table}[!h]
	\begin{center}
		\begin{tabular}{lcccccc}
		\hline
		\hline\\
		Model& $\chi^2_{\nu}$&$ w_0$& $ q$&$\Omega_{dm,0}$&$\epsilon$&$\xi$\\
		\hline
		\hline\\
		Wang-Meng&0.97&$-1.00$&$1.02^{+0.07}_{-0.08}$&$0.24^{+0.01}_{-0.01}$&$0.00^{+0.08}_{-0.08}$&$-$\\
		Dalal&0.97&$-1.00$&$1.03^{+0.11}_{-0.11}$&$0.24^{+0.01}_{-0.01}$&$-$&$3.18^{+0.63}_{-0.50}$\\
		\hline
		\hline
		\end{tabular}
	\end{center}
	\caption{The best fit and the $1\sigma$ errors for one parameters for Wang-Meng and Dalal vacuum decay models.}
	\label{bf_vd}
\end{table}

\section{Final Remarks}

One of the biggest conundrums of our time is to understand the process that makes the Universe to accelerate.  A clue about its dynamics can be given by its total energy distribution and by the sources of energy.  Matter fields, like barionic matter and radiation are clearly sources of energy.  Besides, two different components, the DM and DE are ruling out the dynamical features of the Universe.  It is given to DE the property of being the responsible by cosmic acceleration.   DE has the biggest percentage of occupation in the whole Universe and has a negative pressure.  But its features are not completely discovered and/or understood so far.

Our objective in this work was to analyze some DE models through the point of view of Tsallis nonextensive statistics in order to obtain more clues about its behavior. More specifically, we have used the Verlinde ideas together with Tsallis' formalism to study DE models. As a result we have obtained the nonextensive parameter q close to 1. However, from the errors of the q-parameters measurements shown in tables I and II, the hypothesis of a nonextensive nature for the holographic screen, i.e., the hypothesis that the bits obey the Tsallis nonextensive statistical mechanics in the context of cosmological models is quite plausible.

\begin{acknowledgments}

\ni The research of RCN was supported by CAPES-Brazil.  EMCA would like to thank CNPq (Conselho Nacional de Desenvolvimento Cient\' ifico e Tecnol\'ogico), Brazilian scientific support agency, for partial financial support.

\end{acknowledgments}

\end{document}